\documentclass[11pt]{article}
\pdfoutput=1
\usepackage{jheppub,amsmath,amssymb}
\usepackage{bm}
\usepackage[table]{xcolor}
\usepackage{graphics}
\usepackage{tikz}
\usetikzlibrary{arrows,positioning,decorations.markings,decorations.pathmorphing,calc}
\usepackage{todonotes}

\definecolor{hyperref}{RGB}{026,028,087}
\def\be{\begin{equation}}
\def\ee{\end{equation}}

\newcommand{\mn}{\mu\nu}

\newcommand{\St}{{St\"uckelberg} }

\newcommand{\lambdamin}{\lambda_\text{min}}

\graphicspath{{./}}
\notoc

\begin{document}

\title{Strong-coupling scales and the graph structure of multi-gravity theories}

\author[a]{James H. C. Scargill,}
\author[b]{Johannes Noller}

\affiliation[a]{Theoretical Physics, University of Oxford, DWB, Keble Road, Oxford, OX1 3NP, UK} 
\affiliation[b]{Astrophysics, University of Oxford, DWB, Keble Road, Oxford, OX1 3RH, UK} 

\emailAdd{james.scargill@physics.ox.ac.uk}
\emailAdd{noller@physics.ox.ac.uk}

\abstract{
In this paper we consider how the strong-coupling scale, or perturbative cutoff, in a multi-gravity theory depends upon the presence and structure of interactions between the different fields. This can elegantly be rephrased in terms of the size and structure of the `theory graph' which depicts the interactions in a given theory. We show that the question can be answered in terms of the properties of various graph-theoretical matrices, affording an efficient way to estimate and place bounds on the strong-coupling scale of a given theory. In light of this we also consider the problem of relating a given theory graph to a discretised higher dimensional theory, \`a la dimensional deconstruction.}


\maketitle
\newpage

\section{Introduction} \label{sec-intro}

Multi-gravity theories, or theories of multiple interacting spin-2 fields, have been extensively studied in recent years. For consistency, all but one of the fields must be massive \cite{Boulanger:2000rq}, leading to an intimate connection with massive gravity (for reviews see \cite{Hinterbichler:2011tt, deRham:2014zqa}). For a long time it was thought to be impossible to construct a consistent non-linear theory of massive gravity \cite{Boulware:1973my}, yet this proved to be false \cite{deRham:2010ik,deRham:2010kj,Hassan:2011vm}. 
To write a mass term for the metric requires the introduction of a new, fiducial rank 2 tensor field \cite{deRham:2010kj, Hassan:2011hr, Hassan:2011tf}, which can be rendered dynamical, leading to bi-gravity \cite{Hassan:2011zd, Hassan:2011ea}; multi-gravity is then a generalisation of this, involving further dynamical rank 2 tensor fields, and interactions among them \cite{Hinterbichler:2012cn}.
 
Such theories are necessarily effective field theories which have some cutoff -- an energy scale above which they are not valid -- along with some energy scale at which perturbation theory breaks down, and higher order interaction terms become relevant (the so-called `strong coupling' scale); these need not be the same, and the latter only indicates the energy scale above which the theory is strongly coupled. 
Whether, in the case of multi-gravity theories, the strong coupling scale should also be considered as the cutoff of the effective field theory is a matter of some debate (for a general effective field theory argument to that effect see \cite{Aydemir:2012nz}) since these theories are known to be ghost free up to the Planck scale. This is especially relevant in the context of local (e.g. solar system) gravitational tests if such a modification of gravity is invoked to generate self-accelerating cosmological solutions (and thereby be a step towards solving the cosmological constant problem), since is such cases the strong coupling scale is usually very low, typically on the order of $10^3$ km.\footnote{It is worth noting that this value is environment dependent \cite{Burrage:2012ja,deRham:2014zqa}.}

We do not dwell on this question here, as in any case, above the strong coupling scale the nature of the theory clearly must change.  Instead we focus on how the strong coupling scale itself depends on the nature of the multi-gravity theory, and in particular on the size and structure of the `theory graph' (explained in section \ref{sec-theory graphs}). We will find that the structure of the graph does have a large impact, since for example the strong coupling scale of a theory based on a path graph decreases with the number of gravitons as $\sim 1/\sqrt{N}$, whereas that based on a star graph does not change.

A further aspect which provides motivation for studying these theories is dimensional deconstruction, which seeks to approximate a Kaluza-Klein reduced theory by discretising the extra dimension \cite{ArkaniHamed:2001ca}; multi-gravity theories naturally appear out of such a framework \cite{deRham:2013awa}.

This paper has the following structure. In section \ref{sec-recap} we recapitulate the results for the strong coupling scale in dRGT massive gravity and HR bi-gravity, as well as giving a simple tri-gravity example to show how, and why, this can change as more spin-2 fields are introduced. Section \ref{sec-theory graphs} discusses the graph representation of multi-gravity theories, and derives bounds on the strong coupling scale which depend only on properties of the graph. In section \ref{sec-graviton masses} this is related to dimensional deconstruction, possible interpretation of general multi-gravity theories in terms of discretisation of higher-dimensional theories is discussed, and we briefly touch upon the coupling of these theories to matter. Finally, section \ref{sec-conc} concludes. Throughout this paper we will work in four spacetime dimensions, though an extension to arbitrary dimensions does not essentially change the conclusions.

\section{Strong coupling scales in massive, bi-, and multi-gravity}\label{sec-recap}

To find the strong coupling scale for the theory we employ the \St trick to restore diffeomorphism invariance to all the spin-2 fields and expand around Minkowski space, in order to make manifest the different degrees of freedom (for a review see \cite{Hinterbichler:2011tt}). There will exist operators with dimension greater than four, suppressed by certain scales, the smallest of which gives the strong coupling scale. We will recapitulate this for massive- and bi-gravity, before giving a simple tri-gravity example, which contains a lot of the qualitatively different features of multi-gravity. (See \cite{deRham:2011rn}, \cite{Fasiello:2013woa}, and \cite{Noller:2013yja} respectively for more details.)

\subsection{Massive gravity}

In \emph{massive} gravity there is one dynamical metric, $g$, which transforms under general coordinate transformations ($GC$), and one fiducial non-dynamical one, $f$, which does not transform under $GC$. The action consists of a kinetic term (Ricci scalar) for $g$, and the following dRGT interaction terms:
\be \label{mga}
\mathcal{L} = \mathcal{L}_\text{kin} +\mathcal{L}_\text{int} = M^2 \sqrt{-g}R[g]+ m^2 M^2 \sqrt{-g} \sum_{i=0}^4 \beta_i e_{i}\left( \sqrt{g^{-1} f} \right),
\ee
where in massive gravity $M$ would be the Planck mass for $g$, but we leave it general here as this is not necessarily the case in multi-gravity, and $m$ is a mass parameter which controls the size of the graviton mass. 

The interaction terms break the $GC$ symmetry, which however can be restored by introducing the \St fields $Y^\mu$, in a manner patterned after the gauge symmetry we wish to (re-)introduce:
\be
f_{\mn}(x) \to F_{\mn}(x) = \partial_\mu Y^\alpha \partial_\nu Y^\beta f_{\alpha\beta}(Y(x)).
\ee
The transformation properties of $Y^\mu$ are such that $F_{\mn}$ transforms as a tensor under $GC$. We then expand the metrics around flat space, and the \St field around the identity,
\be
g_{\mn} = \eta_{\mn} + h_{\mn}, \qquad f_{\mn} = \eta_{\mn}, \qquad Y^\mu(x) = x^\mu + A^\mu,
\ee
before finally introducing a $U(1)$ \St field for $A^\mu$: $A^\mu \to A^\mu + \partial^\mu \pi$. 

Note that the \St scalar field, $\pi$, always appears with two derivatives acting on it, and so cannot at this stage have a canonical kinetic term. It gains one through mixing with the spin-2 field $h_{\mn}$: at the quadratic order these fields are mixed, which can be removed via a field redefinition of the form $\sim h_{\mn} \to h_{\mn} - \pi \eta_{\mn}$. 

As well as removing the $h\pi$ mixing and introducing a canonical kinetic term for $\pi$, this field redefinition also introduces terms of the form $\pi (\partial^2 \pi)^n$, which will turn out to be those suppressed by the lowest scale. Once all the fields are canonically normalised one has
\be
{\mathcal L} \supset \left(\frac{1}{4}(\beta_2 + \beta_3) \right) \frac{1}{\Lambda_3^3} \pi X^\mu_{(2),\mu} (\pi) + \left( \frac{1}{6} \beta_3 \right) \frac{1}{\Lambda_3^6} \pi X^\mu_{(3),\mu} (\pi),
\ee
and hence the strong coupling scale of dRGT massive gravity is $\Lambda_3 = \left( m^2 M \right)^{1/3}$.

\subsection{Bi-gravity}

In bi-gravity the action is as in massive-gravity, with the addition of a kinetic term for $f$, which is now dynamical. The two dynamical fields each transform under a different copy of $GC$; the interaction term breaks the symmetry $GC_g \times GC_f$ down to the diagonal subgroup. 
\be
\mathcal{L} =  M^2 \sqrt{-g}R[g] +  M^2 \sqrt{-f}R[f] +\frac{1}{2}\mathcal{L}_\text{int},
\ee
where $\mathcal{L}_\text{int}$ is as in \eqref{mga}\footnote{Note that since $\sqrt{-g} e_m \left( \sqrt{g^{-1} f} \right) = \sqrt{-f} e_{D-m} \left( \sqrt{f^{-1} g} \right)$, ${\mathcal L}_{\rm int}$ does not preference either of the two fields.}, the factor of $1/2$ comes from replacing $M$ in the interaction term by $M_{\rm eff} = (M_g^{-2} + M_f^{-2})^{-1/2}$
and we have chosen the `Planck masses' $M_g$ and $M_f$ of the two fields to be the same.

Compared with massive gravity there is now a choice as to how we introduce the \St fields: we can again make $f$ transform as a tensor under $GC_g$ (and be invariant under $GC_f$) via $f_{\mn}(x) \to F_{\mn}(x) = \partial_\mu Y_{f \to g}^\alpha \partial_\nu Y_{f \to g}^\beta f_{\alpha\beta}(Y_{f \to g}(x))$, or we can do the opposite, and change the transformation properties of $g$ (making it a tensor under $GC_f$ and invariant under $GC_g$) via $g_{\mn}(x) \to G_{\mn}(x) = \partial_\mu Y_{g \to f}^\alpha \partial_\nu Y_{g \to f}^\beta g_{\alpha\beta}(Y_{g \to f}(x))$. Comparing the two cases, one finds $Y_{g \to f} = Y_{f \to g}^{-1}$, and that the \St scalar fields in each case,
\be
Y^{\mu}_{f \to g}(x) = x^\mu + \partial^\mu \pi, \qquad Y^\mu_{g \to f}(x) = x^\mu + \partial^\mu \phi,
\ee
are related to that in the other case by the so-called Galileon duality \cite{deRham:2013hsa}, which in itself is a direct consequence of the different, yet physically equivalent, ways of introducing \St fields.

 The terms suppressed by $\Lambda_3$ are now
\be
\left(\frac{1}{12}( 5\beta_2 + 4\beta_3 - \beta_4) \right) \frac{1}{\Lambda_3^3} \pi X^\mu_{(2),\mu} (\pi) + \left( \frac{1}{24} (7\beta_3 - \beta_4) \right) \frac{1}{\Lambda_3^6} \pi X^\mu_{(3),\mu} (\pi),
\ee
where terms involving $\phi$ have been rewritten using the Galileon duality relation $\phi = -\pi + \frac{1}{2} \left( \partial \pi \right)^2 + \cdots$ \cite{Fasiello:2013woa,Noller:2015eda}. Thus the strong coupling scale is still effectively $\Lambda_3$.

\subsection{Tri-gravity}

Consider now a tri-metric theory in which two of the fields (labelled 2 and 3) interact only with a third (labelled 1), and whose interaction Lagrangian is (also see figure \ref{fig-P_3})
\be
\mathcal{L}_\text{int} = m^2 M^2 \left[ \sqrt{-g_{(2)}} \sum_i \beta^{(1,2)}_i e_{i}\left( \sqrt{g_{(2)}^{-1} g_{(1)}} \right) + \sqrt{-g_{(3)}} \sum_i \beta^{(1,3)}_i e_{i}\left( \sqrt{g_{(3)}^{-1} g_{(1)}} \right) \right], \label{trimetric lagrangian}
\ee
and the kinetic sector for this and all subsequent $N$-metric theories consists of a sum of the corresponding $N$ Einstein-Hilbert terms, just as in the bigravity example above.

We now introduce \St fields in each term mapping $g_{(1)}$ to site 2 or 3 as appropriate.\footnote{As explained in more detail in \cite{Noller:2013yja}, when applying the \St trick to a multi-gravity theory, one can treat each interaction term independently, and for a given interaction term $f(g^{(1)},g^{(2)})$ one can choose to make $g^{(2)}$ transform under the gauge symmetry of $g^{(1)}$, or vice versa. The two possibilities are equivalent and related by the so-called Galileon duality \cite{deRham:2013hsa, Noller:2015eda}.}  For the remainder of this subsection we set $m = M = 1$ in order to focus on the interaction structure of the theory, since the explicit mass scales are the same as in bigravity. To cubic order in the fields the scalar-tensor mixing then looks like
\begin{align}
\mathcal{L}_{h\pi} = &h^{(1)}_{\mn} \left[ a_1 X_{(1)}^{\mn}(\phi^{(1,2)}) + b_{1,R} X_{(2)}^{\mn}(\phi^{(1,2)}) + a_2 X_{(1)}^{\mn}(\phi^{(1,3)}) + b_{2,R} X_{(2)}^{\mn}(\phi^{(1,3)}) \right] \nonumber \\
&+ h^{(2)}_{\mn} \left[ a_1 X_{(1)}^{\mn}(\pi^{(1,2)}) + b_{1,L} X_{(2)}^{\mn}(\pi^{(1,2)})  \right] + h^{(3)}_{\mn} \left[ a_2 X_{(1)}^{\mn}(\pi^{(1,3)}) + b_{2,L} X_{(2)}^{\mn}(\pi^{(1,3)})  \right], \label{scalar-tensor mixing}
\end{align}
where the coefficients $a_i, b_{i,R}, b_{i,L}$ are functions of the $\beta_i^{(j,k)}$ parameters (and the $L,R$ labels refer to the direction of the \St fields). One can de-mix this through the following field redefinition:
\be
h^{(1)}_{\mn} \to h^{(1)}_{\mn} - (a_1 \phi^{(1,2)} + a_2 \phi^{(1,3)} ) \eta_{\mn}, \qquad h^{(i)}_{\mn} \to h^{(i)}_{\mn} - a_{i-1} \pi^{(1,i)} \eta_{\mn} \quad \text{for i = 2,3}.
\ee
To examine the quadratic and cubic order terms we must pick one from each pair of a \St field and its dual, and rewrite the other in terms of that via
\be
\phi = -\pi + \frac{1}{2} \left( \partial \pi \right)^2 + \cdots,
\ee
(or equivalently for $\pi$ in terms of $\phi$). Calling the so-chosen fields $\rho_{(i)}$, $i = 1,2$, and writing $\tilde{\rho}_{(i)} = a_{i-1} \rho_{(i)}$, one has the following kinetic terms
\be
\mathcal{L}_{\rho\Box\rho} \propto
\begin{pmatrix}
\tilde{\rho}_{(1)} & \tilde{\rho}_{(2)} 
\end{pmatrix}
\begin{pmatrix}
2 & \sigma \\
\sigma & 2
\end{pmatrix}
\begin{pmatrix}
\Box \tilde{\rho}_{(1)} \\
\Box \tilde{\rho}_{(2)}
\end{pmatrix}
= \tilde{\rho}^T K \Box \tilde{\rho}, \label{tri-metric K}
\ee
where $\sigma = 1$ if $\{ \rho_{(i)} \} = \{\pi^{(1,2)}, \pi^{(1,3)} \} \text{ or } \{\phi^{(1,2)}, \phi^{(1,3)} \}$, i.e. either $g_{(1)}$ is being always mapped to the other sites, or $g_{(2),(3)}$ are both being mapped to site 1; and $\sigma = -1$ otherwise, i.e. $g_{(1)}$ is mapped to another site in one interaction term, but not in the other, which corresponds to $\{ \rho_{(i)} \} = \{\pi^{(1,2)}, \phi^{(1,3)} \} \text{ or } \{\phi^{(1,2)}, \pi^{(1,3)} \}$. See figure \ref{fig-mapping directions}.

At cubic order there are two types of terms: those arising from $\rho_{(i)} X^\mu_{(2)\mu}(\rho_{(j)})$, as appear in massive gravity, and those arising from the second order expansion of one of the fields in a nominally quadratic term, e.g. $\phi^{(1,2)} \Box \phi^{(1,3)} \to \frac{1}{2} (\partial \rho_{(1)})^2 \Box \rho_{(2)}$. We call these respectively, A-, and B-type terms, and write
\be
\mathcal{L}_A \propto \begin{pmatrix}
\tilde{\rho}_{(1)} & \tilde{\rho}_{(2)} 
\end{pmatrix}
C^A 
\begin{pmatrix}
X^\mu_{(2)\mu}(\tilde{\rho}_{(1)}) \\
X^\mu_{(2)\mu}(\tilde{\rho}_{(2)})
\end{pmatrix}
\quad \text{and} \quad
\mathcal{L}_B \propto \begin{pmatrix}
\Box \tilde{\rho}_{(1)} & \Box \tilde{\rho}_{(2)}
\end{pmatrix}
C^B
\begin{pmatrix}
(\partial\tilde{\rho}_{(1)})^2 \\
(\partial\tilde{\rho}_{(2)})^2
\end{pmatrix},
\ee
where the coefficient matrices $C^{A,B}$ depend on how the links are oriented, and are displayed in table \ref{tab-trimetric Cs}.

\begin{table}[ht]
\centering
\begin{tabular}{c | c c}
$\{ \rho_{(1)}, \rho_{(2)} \}$ & $C^A$ & $C^B$ \\
\hline
$\{ \pi^{(1,2)}, \pi^{(1,3)} \}$ & 
$\begin{pmatrix}
 a_1^{-2}(b_{1,L} - b_{1,R}) & - a_2^{-2} b_{2,R} \\
- a_1^{-2} b_{1,R} & a_2^{-2} (b_{2,L} - b_{2,R})
\end{pmatrix}$ & 
$\begin{pmatrix}
a_1^{-1} & a_2^{-1} \\
a_1^{-1} & a_2^{-1}
\end{pmatrix}$
\\
$\{ \phi^{(1,2)}, \phi^{(1,3)} \}$ & $
\begin{pmatrix}
a_1^{-2} (b_{1,R} - b_{1,L}) & a_2^{-2} b_{2,R} \\
a_1^{-2} b_{1,R} & a_2^{-2} (b_{2,R} - b_{2,L})
\end{pmatrix}$ & 
$\begin{pmatrix}
a_1^{-1} & 0 \\
0 & a_2^{-1}
\end{pmatrix}$ \\
$\{ \pi^{(1,2)}, \phi^{(1,3)} \}$ & $
\begin{pmatrix}
a_1^{-2} (b_{1,L} - b_{1,R}) & -a_2^{-2} b_{2,R} \\
a_1^{-2} b_{1,R} & a_2^{-2} (b_{2,R} - b_{2,L})
\end{pmatrix}$ & 
$\begin{pmatrix}
a_1^{-1} & 0 \\
-a_1^{-1} & a_2^{-1}
\end{pmatrix}$ \\
$\{ \phi^{(1,2)}, \pi^{(1,3)} \}$ & $
\begin{pmatrix}
a_1^{-2} (b_{1,R} - b_{1,L}) & a_2^{-2} b_{2,R} \\
- a_1^{-2} b_{1,R} & a_2^{-2} (b_{2,L} - b_{2,R})
\end{pmatrix}$ & 
$\begin{pmatrix}
a_1^{-1} & - a_2^{-1} \\
0 & a_2^{-1}
\end{pmatrix}$
\end{tabular}
\caption{The matrices of coefficients of cubic terms (before quadratic demixing and canonical normalisation of the modes), for the different possible orientations of the links.} \label{tab-trimetric Cs}
\end{table}

As the scalars are mixed at the quadratic level, in order to ask questions concerning when interactions become strongly coupled we must find the actual propagating modes, i.e. we must diagonalise the kinetic matrix $K$.\footnote{We also need to diagonalise their mass matrix, however in practice this does not have a significant impact on the strong coupling scale, and so we ignore it here, but comment further in section \ref{sec-mass}.} Transforming to the eigenbasis of $K$ involves $\tilde{\rho} \to U \chi$, where $U$ is the matrix whose columns are the normalised eigenvectors of $K$; the kinetic terms will then look like $\sum_{i=1}^2 \lambda_i \chi^{(i)} \Box \chi^{(i)}$, and so finally we must canonically normalise the fields, $\chi^{(i)} \to \chi^{(i)} / \sqrt{\lambda_i}$.

Once this is done the cubic terms will look like
\be
\mathcal{L}_A = \sum_{ijk} \frac{\tilde{C}^A_{ijk}}{\Lambda_3^3} \chi^i X^\mu_{(2),\mu} \left( \chi^j, \chi^k \right), \quad \text{and} \quad \mathcal{L}_B = \sum_{ijk} \frac{\tilde{C}^B_{ijk}}{\Lambda_3^3} \Box \chi^i \partial_\mu \chi^j  \partial^\mu \chi^k,
\ee
where 
\be
\tilde{C}_{ijk} = \frac{1}{\sqrt{\lambda_i \lambda_j \lambda_k}} \sum_{l,m} C_{l m} U_{l i} U_{m j} U_{m k}, \label{Ctilde}
\ee
and the explicit mass scale dependence via $\Lambda_3^3$ has been reintroduced. 
The least suppressed cubic interaction is the one with the largest (in magnitude) element of $\tilde{C}$, $\tilde{C}_\text{max}$ and hence the effective strong coupling scale is $\Lambda_3 / \left( \tilde{C}_\text{max} \right)^{1/3}$.

The strong coupling scale is thus shifted from the massive-/bi-gravity value. Whilst in the tri-metric case, assuming that the coefficients in front of the initial interaction terms are of the same order, this shift is not terribly large,\footnote{Though see \cite{Noller:2013yja} for examples of how this is no longer true when the assumption is invalid.} when the number of fields involved is large the effect can be significant, as we show below. The reader may wonder whether higher order terms (quartic in the scalar and above) could lead to an effective strong coupling scale which is lower than this. For tri-gravity this may be possible, since the numbers are still $\mathcal{O}(1)$, however as we shall explain in section \ref{sec-higherorder}, one would expect higher order terms to be less suppressed than cubic for larger numbers of interacting spin-2 fields.

We emphasise that this requirement to de-mix the scalar fields, and the resulting shift in the strong coupling scale, is a qualitatively different feature of multi-gravity, which does not appear in bi- or massive- gravity.

\section{Theory graphs}\label{sec-theory graphs}

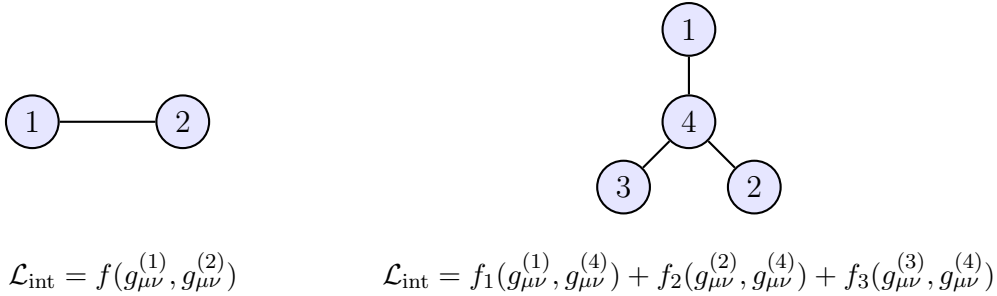
\begin{figure}[tp]
\centering
\begin{tikzpicture}[-,>=stealth',shorten >=0pt,auto,node distance=2cm,
  thick,main node/.style={circle,fill=blue!10,draw,font=\sffamily\large\bfseries},arrow line/.style={thick,-},barrow line/.style={thick,->},no node/.style={plain},rect node/.style={rectangle,fill=blue!10,draw,font=\sffamily\large\bfseries},red node/.style={rectangle,fill=red!10,draw,font=\sffamily\large\bfseries},green node/.style={circle,fill=green!20,draw,font=\sffamily\large\bfseries},yellow node/.style={rectangle,fill=yellow!20,draw,font=\sffamily\large\bfseries}]

	\node[main node](100){$1$};
	\node[main node](101)[right of=100]{$2$};
	\node[draw=none,fill=none](83)[right=0.7cm of 100]{};
    
	\node[main node](200)[right=6cm of 101]{$4$};
	\node[main node](201)[above=0.5cm of 200]{$1$};
	\node[main node](202)[below right=0.5cm of 200]{$2$};
	\node[main node](203)[below left=0.5cm of 200]{$3$};

	\path[every node/.style={font=\sffamily\small}]
	(100) edge node {} (101)
	(200) edge node {} (201)
	(200) edge node {} (202)
	(200) edge node {} (203);

	\node[draw=none,fill=none](1)[below of=83]{$\mathcal{L}_\text{int} = f(g^{(1)}_{\mn}, g^{(2)}_{\mn})$};
   
	\node[draw=none,fill=none](2)[below of=200]{$\mathcal{L}_\text{int} = f_1(g^{(1)}_{\mn}, g^{(4)}_{\mn})+ f_2(g^{(2)}_{\mn}, g^{(4)}_{\mn}) + f_3(g^{(3)}_{\mn}, g^{(4)}_{\mn})$};
   
\end{tikzpicture}

\caption{Examples of different types of theory graphs and their corresponding interaction Lagrangians; vertices correspond to metrics, and when two metrics are present in the same interaction term the corresponding vertices are connected by an edge.} \label{fig-theory graph examples}
\end{figure}

The structure of a multi-gravity theory can be conveniently encoded in a so-called theory graph \cite{ArkaniHamed:2002sp,Hinterbichler:2012cn,Noller:2013yja}. Each spin-2 field in the theory is represented by a vertex, and where two fields appear in an interaction term in the action the corresponding vertices in the theory graph are connected. Examples are given in figure \ref{fig-theory graph examples}. Note that we are only considering superpositions of bigravity-like interaction terms, but in general one may also write interaction terms which involve more than two fields \cite{Hinterbichler:2012cn}. Also, certain aspects of multi-gravity theories, in particular the cosmology, have been studied from a graph theoretical perspective in \cite{Hanada:2010js}.

Already some structural properties of the theory graph are known to have a bearing on the theory. In particular, if the theory graph contains cycles, then the equivalence between the metric and vielbein versions of the theory breaks down \cite{Hinterbichler:2012cn}, and in \cite{Scargill:2014wya} it was shown that a cycle causes a ghost to appear at a scale lower than $\Lambda_3$ in the metric version of the theory, which lowers the strong coupling scale and the cutoff. Furthermore, the vielbein version of the theory, when examined in the decoupling limit, which makes explicit the strong coupling scale, becomes exceedingly complicated when there is a cycle in the graph \cite{Scargill:2014wya}. 

For these reasons, in what follows we will focus on theory graphs which are acyclic (i.e. they are tree graphs), and it does not matter which version (metric or vielbein) of the theory we use; for concreteness we use the metric version.

Also note that to encode the full information about the theory this would be a weighted graph, with the weight of each edge depending on the coefficient of the corresponding interaction term, however in order to focus on the effect of the structure of the graph we will choose these coefficients so that the we can consider simply an unweighted graph.

\subsection{Relation between the theory graph and the strong coupling scale} \label{sec-theory graph to strong coupling}

As explained in section \ref{sec-recap}, the strong coupling scale of a multi-gravity theory (in which all the interaction mass scales are identical) is not simply $\Lambda_3$ because one must de-mix and canonically normalise the \St scalar fields, which leads to a hierarchy in the values of the coefficients in front of the non-linear interaction terms (i.e. the Wilson coefficients). From \eqref{Ctilde} we see that the eigenvalues of the \St scalar kinetic matrix for the de-mixed modes in question are a key ingredient in the size of the coefficient. Thus a first step in finding the strong coupling scale is to find the smallest eigenvalue of the kinetic matrix.

\begin{figure}[tp]
\centering
\begin{tikzpicture}[-,>=stealth',shorten >=0pt,auto,node distance=2cm,
  thick,main node/.style={circle,fill=blue!10,draw,font=\sffamily\large\bfseries},arrow line/.style={thick,-},barrow line/.style={thick,->},no node/.style={plain},rect node/.style={rectangle,fill=blue!10,draw,font=\sffamily\large\bfseries},red node/.style={rectangle,fill=red!10,draw,font=\sffamily\large\bfseries},green node/.style={circle,fill=green!20,draw,font=\sffamily\large\bfseries},yellow node/.style={rectangle,fill=yellow!20,draw,font=\sffamily\large\bfseries}]

	\node[main node](100){$2$};
	\node[main node](101)[right of=100]{$1$};
	\node[main node](102)[right of=101]{$3$};

	\path[every node/.style={font=\sffamily\small}]
	(100) edge node {} (101)
	(101) edge node {} (102);
   
\end{tikzpicture}

\caption{The path graph on three vertices, $P_3$.} \label{fig-P_3}
\end{figure}
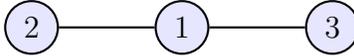

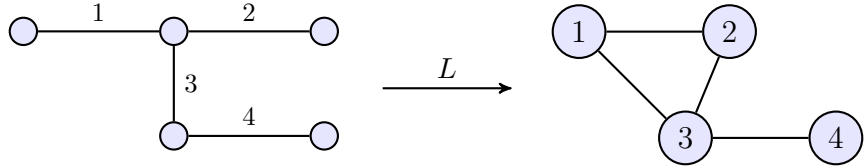
\begin{figure}[tp]
\centering
\begin{tikzpicture}[-,>=stealth',shorten >=0pt,auto,node distance=2cm,
  thick,main node/.style={circle,fill=blue!10,draw,font=\sffamily\large\bfseries},arrow line/.style={thick,-},barrow line/.style={thick,->},no node/.style={plain},rect node/.style={rectangle,fill=blue!10,draw,font=\sffamily\large\bfseries},red node/.style={rectangle,fill=red!10,draw,font=\sffamily\large\bfseries},green node/.style={circle,fill=green!20,draw,font=\sffamily\large\bfseries},yellow node/.style={rectangle,fill=yellow!20,draw,font=\sffamily\large\bfseries}]

	\node[main node](100){};
	\node[main node](101)[right of=100]{};
	\node[main node](102)[right of=101]{};
	\node[main node](103)[below=1cm of 101]{};
	\node[main node](104)[right of=103]{};

	\node[draw=none,fill=none](150)[above right=0.5cm of 104]{};
	\node[draw=none,fill=none](151)[right of=150]{};

	\node[main node](200)[above right=0.5cm of 151]{$1$};
	\node[main node](201)[right of=200]{$2$};
	\node[main node](202)[below right of=200]{$3$};
	\node[main node](203)[right of=202]{$4$};

	\path[every node/.style={font=\sffamily\small}]
	(100) edge node {$1$} (101)
	(101) edge node {$2$} (102)
	(101) edge node {$3$} (103)
	(103) edge node {$4$} (104);

	\path[every node/.style={font=\sffamily\small}]
	(200) edge node {} (201)
	(201) edge node {} (202)
	(200) edge node {} (202)
	(202) edge node {} (203);

	\draw[->] (150) -- (151) node[midway,above] {$L$};

\end{tikzpicture}

\caption{The operation of the line graph operator, $L$.} \label{fig-L example}
\end{figure}

In order to see how this is related to the theory graph let us return to the tri-metric theory considered in the previous section, whose theory graph is simply the path graph on three vertices,\footnote{Note that as we are considering only acyclic graphs, this is the only possibility with three vertices (up to vertex relabelling).} $P_3$, as shown in figure \ref{fig-P_3}. If we choose the direction of the \St mapping so that $\sigma = 1$, we see from \eqref{tri-metric K} that (in matrix notation) $K = 2 I_2 + A(P_2)$, where $I_N$ is the $N$-dimensional identity matrix and $A(G)$ is the \emph{adjacency matrix} of a graph $G$, which is defined as
\be
A(G)_{ij} = 
\begin{cases}
1 & \text{if vertices }i\text{ and }j\text{ are joined by an edge}, \\
0 & \text{else};
\end{cases}
\ee
we also note that $P_2$ is the \emph{line graph} of $P_3$, where the line graph, $L_G$, is the graph whose vertices are the edges of $G$, and two vertices in $L_G$ are joined by an edge if the corresponding edges in $G$ are connected to the same vertex. This is illustrated by an example in figure \ref{fig-L example}.

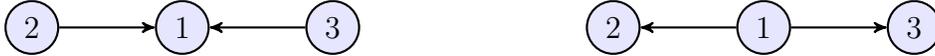
\begin{figure}[tp]
\centering
\begin{tikzpicture}[->,>=stealth',shorten >=0pt,auto,node distance=2cm,
  thick,main node/.style={circle,fill=blue!10,draw,font=\sffamily\large\bfseries},arrow line/.style={thick,-},barrow line/.style={thick,->},no node/.style={plain},rect node/.style={rectangle,fill=blue!10,draw,font=\sffamily\large\bfseries},red node/.style={rectangle,fill=red!10,draw,font=\sffamily\large\bfseries},green node/.style={circle,fill=green!20,draw,font=\sffamily\large\bfseries},yellow node/.style={rectangle,fill=yellow!20,draw,font=\sffamily\large\bfseries}]

	\node[main node](100){$2$};
	\node[main node](101)[right of=100]{$1$};
	\node[main node](102)[right of=101]{$3$};

	\node[main node](200)[right=3cm of 102]{$2$};
	\node[main node](201)[right of=200]{$1$};
	\node[main node](202)[right of=201]{$3$};

	\path[every node/.style={font=\sffamily\small}]
	(100) edge node {} (101)
	(102) edge node {} (101)
	(201) edge node {} (200)
	(201) edge node {} (202);
   
\end{tikzpicture}

\caption{\St mapping directions allowed by $\sigma = 1$ in \eqref{tri-metric K}.} \label{fig-mapping directions}
\end{figure}

The choice $\sigma = 1$ corresponds to choosing the direction of the \St mapping such that either $g^{(2)}$ and $g^{(3)}$ are both mapped to the transformation properties of $g^{(1)}$ in their respective interaction terms, \emph{or} $g^{(1)}$ is mapped to the transformation properties of $g^{(2)}$ or $g^{(3)}$ in the respective interaction terms. Or in other words, that each vertex in the theory graph either only has fields being mapped to it, or away from it, as shown in figure \ref{fig-mapping directions}. 

For a general multi-gravity theory we can consider all \St scalars in this way, and furthermore for a tree graph we can always orient the edges so that each vertex has only inward or only outward edges (corresponding to $\sigma = 1$; on the other hand, choosing $\sigma = -1$ cannot be consistently extended to graphs with vertices of degree higher than two). Thus for a general theory graph $T$ with $N$ vertices\footnote{And since this is a tree graph it will have $N-1$ edges.} we have
\be \label{KT}
K(T) \propto 2 I_{N-1} + A(L_T).
\ee

Thus the problem of finding the smallest eigenvalue of the kinetic matrix is the same as finding the smallest eigenvalue of the line graph. As we shall later see, $\lambdamin(K(T))$ is equal to the \emph{algebraic connectivity} of the graph $T$. The following subsection is devoted to placing bounds on the smallest eigenvalue of the line graph of a tree graph.

\subsection{Placing bounds on the smallest eigenvalue of $A(L_T)$}

Some bounds are already known about the smallest eigenvalue of the line graph of a tree; in particular from \cite{Yan} we have
\be
-2 \cos \left( \frac{\pi}{N} \right) \leq \lambdamin(L_T) \leq -1, \label{bounds1}
\ee
where the lower bound is saturated iff $T = P_N$, and the upper bound iff $T = S_N$, i.e. the `star graph', which consists of one vertex connected to $N-1$ further vertices which are not connected amongst themselves (the right graph in figure \ref{fig-theory graph examples} is $S_4$, for example).\footnote{In standard notation the star graph would be $K_{1,N-1}$, but we call it $S_N$ here to avoid confusion with the kinetic matrix $K$.}  There is quite a gulf between these two graphs and their corresponding bounds, and in fact the former results in a strong coupling scale which goes $\sim 1/\sqrt{N}$ whereas the latter in one which doesn't scale with $N$. This makes it difficult to draw conclusions about the strong coupling scale for a general tree. Thus we attempt to find bounds which are tighter, though note that they must necessarily depend on further properties of the graph, as the bounds \eqref{bounds1} are saturated in certain circumstances.

Incidentally we point out that from \eqref{bounds1} we have $\lambdamin > -2$ (for any finite $N$) and hence $K$ is positive definite (cf. equation \eqref{KT}), which agrees with expectations since we know that in the absence of cycles all the \St scalar modes have a kinetic term, all of which are of the right sign.\footnote{Extending this to graphs which contain cycles one has that, unless G is a tree or contains at least one odd cycle, $\lambdamin(L_G) = -2$ (and $\lambdamin(L_G) \geq -2$ in general) \cite{Doob197340}, which means that not all the \St scalar modes have a kinetic term -- leading to all sorts of problems (for details see \cite{Scargill:2014wya}). One might be tempted to conclude that this means that theory graphs containing at least one odd cycle can avoid the problems, but note that our prescription concerning the directions of \St mapping fails in the presence of an odd cycle and we no longer have $K \propto 2 I + A(L_G)$.}

\subsubsection{Upper bound}

Consider adding a vertex (and edge) to $T$ to form a new tree $T'$. By suitably relabelling vertices we have
\be
A(L_{T'}) =
\begin{pmatrix}
A(L_T) & a\\
a^T & 0
\end{pmatrix},
\ee
where $a^T = (0, \ldots, 0, \overbrace{1, \ldots, 1}^m)$, and $m$ is the degree of the vertex in $T$ to which the new vertex is attached. Let $v$ be the eigenvector of $A(L_T)$ with the smallest eigenvalue $\lambdamin(L_T)$, and let ${v'}^T = (v^T, 0)$, then by Rayleigh's theorem we have
\be
\lambdamin(L_{T'}) \leq \frac{{v'}^T A(L_{T'}) v'}{{v'}^T v'} = \lambdamin(L_T).
\ee
Thus we see that as a tree grows, the minimum eigenvalue of its line graph can never increase. As we shall see later, this means that we can never raise the strong coupling scale of a theory by adding new vertices and/or edges in this way. 

The \emph{diameter} of a graph, $d(G)$, is defined as the maximum of the set of minimum distances between all pairs of vertices; thus for example $d(P_N) = N-1$ and $d({S_N})$ = 2. Any tree can then be considered to be $P_{d(T)+1}$ with certain `extrusions' attached to it, and thus we immediately have
\be
\lambdamin(L_T) \leq \lambdamin(L_{P_{d(T)+1}}) = -2 \cos \left(\frac{\pi}{d(T)+1} \right).
\ee

\subsubsection{Lower bound}

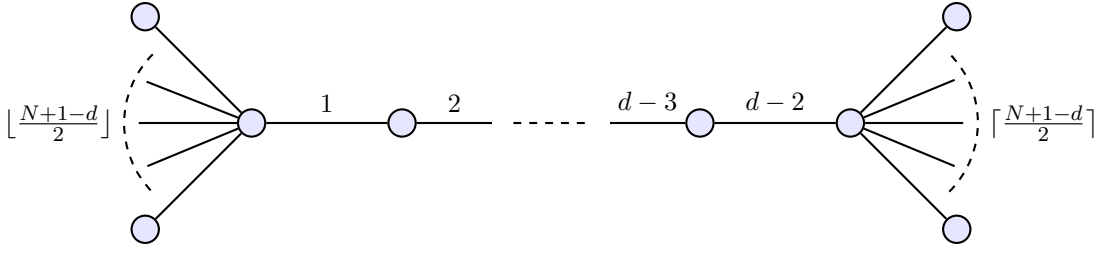
\begin{figure}[tp]
\centering
\begin{tikzpicture}[-,>=stealth',shorten >=0pt,auto,node distance=2cm,
  thick,main node/.style={circle,fill=blue!10,draw,font=\sffamily\large\bfseries},arrow line/.style={thick,-},barrow line/.style={thick,->},no node/.style={plain},rect node/.style={rectangle,fill=blue!10,draw,font=\sffamily\large\bfseries},red node/.style={rectangle,fill=red!10,draw,font=\sffamily\large\bfseries},green node/.style={circle,fill=green!20,draw,font=\sffamily\large\bfseries},yellow node/.style={rectangle,fill=yellow!20,draw,font=\sffamily\large\bfseries}]

	\node[main node](100){};
	\node[main node](101)[right of=100]{};
	\node[draw=none,fill=none](102)[right=1cm of 101]{};
	\node[draw=none,fill=none](103)[right=1cm of 102]{};
	\node[main node](104)[right=1cm  of 103]{};
	\node[main node](105)[right of=104]{};

	\node[main node](99)[above left of=100]{};
	\node[main node](98)[below left of=100]{};

	\node[main node](106)[above right of=105]{};
	\node[main node](107)[below right of=105]{};

	\path[every node/.style={font=\sffamily\small}]
	(100) edge node {$1$} (101)
	(101) edge node {$2$} (102)
	(103) edge node {$d-3$} (104)
	(104) edge node {$d-2$} (105)
	(100) edge node {} (99)
	(100) edge node {} (98)
	(105) edge node {} (106)
	(105) edge node {} (107);

	\draw[-,dashed] (102) to (103);

	\draw[-,dashed] ($(99)+(0.1,-0.5)$) arc (135:225:1.3);
	\draw(100) -- (158.5:1.5);
	\draw(100) -- (180:1.5);
	\draw(100) -- (202.5:1.5);

   	\draw[-,dashed] ($(107)+(-0.1,0.5)$) arc (-45:45:1.3);
	\draw(105) -- ($(105)+(22.5:1.5)$);
	\draw(105) -- ($(105)+(0:1.5)$);
	\draw(105) -- ($(105)+(-22.5:1.5)$);

	\node[draw=none,fill=none](1)[left=1.5cm of 100]{$\lfloor \frac{N+1-d}{2} \rfloor$};
	\node[draw=none,fill=none](2)[right=1.5cm of 105]{$\lceil \frac{N+1-d}{2} \rceil$};

\end{tikzpicture}

\caption{The tree which minimises $\lambdamin(L_T)$ for a given number of edges $N-1$ and diameter $d$. $\lfloor X \rfloor$ denotes the largest integer less than or equal to $X$ and vice versa for $\lceil X \rceil$.} \label{fig-lower bound graph}
\end{figure}

The lower bound of \eqref{bounds1} can also be improved. Given $N-1$ edges and a diameter $d$, the tree whose line graph has the smallest eigenvalue is shown in figure \ref{fig-lower bound graph}. 

Thus the smallest eigenvalue of the associated line graph will give a lower bound for a given diameter and number of edges. The derivation of the characteristic polynomial of the adjacency matrix of the line graph of figure \ref{fig-lower bound graph} can be found in appendix \ref{app-dumbbell graph}, and the result is that the smallest eigenvalue is $-2\cos \theta^*_{N,d}$, where $\theta^*_{N,d}$ is the smallest non-zero root of
\begin{align}
f_{N,d}(\theta)=&\sin((d+1)\theta) + (N-1-d) \sin(d\theta) + \left( \frac{1}{4}(N-3-d)^2 -1 \right) \sin((d-1)\theta) \nonumber \\
&- \frac{1}{2}(N-1-d)^2 \sin((d-2)\theta) + \frac{1}{4}(N-1-d)^2 \sin((d-3)\theta). \label{fNd}
\end{align}
Expanding this about $\theta = 0$ ($\lambda = -2$) we can get a lower bound (valid for $d>2$):\footnote{When $d=2$ the tree must be a star, and we have $\lambdamin(K({S_N})) = 1$.}
\be
\theta^*_{N,d} = \frac{2}{\sqrt{N (d-2)}} \left[1+ \mathcal{O} \left( \frac{1}{Nd}, \left(\frac{d}{N}\right)^2 \right) \right].
\ee

We now have an improved set of bounds for the smallest eigenvalue of the line graph of any tree of given $N,d$:\footnote{These bounds do require knowledge of the diameter of the tree, however in general this is much easier to compute than the spectrum ($\mathcal{O}(N)$ vs. $\mathcal{O}(N^3)$).}
\be
-2 \cos \left(\frac{\pi}{d+1} \right) \geq \lambdamin(L_T) \geq -2 \cos \theta^*_{N,d},
\ee
where the lower bound is saturated by the graph in figure \ref{fig-lower bound graph}. For large $N, d$ this leads to the following bounds on the smallest eigenvalue of the kinetic matrix
\be
\frac{4}{N(d-2)} < \lambdamin(K(T)) < \frac{\pi^2}{(d+1)^2}. \label{lambda bounds}
\ee

\subsection{Bounds on the strong coupling scale} \label{sec-bounds}

To get to the strong coupling scale however requires more than just the eigenvalues of the kinetic matrix, we also need to deal with the sum in \eqref{Ctilde}. From table \ref{tab-trimetric Cs} we see that the form of $C^{A,B}$ depends not just on the relative orientation of the links (related to the choice of \St fields) connected to a given vertex (as for $K$), but also whether those links are oriented towards or away from the vertex. Also relevant are the coefficients in front of the different interaction terms, and thus to avoid unnecessary complication, and to focus on the effect of the structure of the graph, in the sequel we demand that the interaction Lagrangian of each link is of identical form, \emph{with the direction of the links in mind}. 

Let us illustrate what we mean in this last point by an example: in the trimetric theory considered previously, in \eqref{trimetric lagrangian}, since the directions of the links to site 1 are the same, $\beta^{(1,2)}_i = \beta^{(1,3)}_i$; if we were to add another site, connected to site 3, the interaction term would have to be $\sqrt{-g_{(3)}} \sum_i \beta_i e_i \left( \sqrt{g_{(3)}^{-1} g_{(4)}} \right)$, whereas if it was connected to site 1, the interaction term would have to be $\sqrt{-g_{(4)}} \sum_i \beta_i e_i \left( \sqrt{g_{(4)}^{-1} g_{(1)}} \right)$.

\begin{figure}[tp]
\centering
\begin{tikzpicture}[-,>=stealth',shorten >=0pt,auto,node distance=2cm,
  thick,main node/.style={circle,fill=blue!10,draw,font=\sffamily\large\bfseries},arrow line/.style={thick,-},barrow line/.style={thick,->},no node/.style={plain},rect node/.style={rectangle,fill=blue!10,draw,font=\sffamily\large\bfseries},red node/.style={rectangle,fill=red!10,draw,font=\sffamily\large\bfseries},green node/.style={circle,fill=green!20,draw,font=\sffamily\large\bfseries},yellow node/.style={rectangle,fill=yellow!20,draw,font=\sffamily\large\bfseries}]

	\node[main node](100){};
	\node[main node](101)[right of=100]{};
	\node[main node](102)[right of=101]{};
	\node[main node](103)[below=1cm of 101]{};
	\node[main node](104)[right of=103]{};

	\node[draw=none,fill=none](150)[above right=0.5cm of 104]{};
	\node[draw=none,fill=none](151)[above right of=150]{};

	\node[main node](200)[above right=0.5cm of 151]{$1$};
	\node[main node](201)[right of=200]{$2$};
	\node[main node](202)[below right of=200]{$3$};
	\node[main node](203)[right of=202]{$4$};

	\node[draw=none,fill=none](251)[below right of=150]{};

	\node[main node](300)[right=0.5cm of 251]{$1$};
	\node[main node](301)[right of=300]{$2$};
	\node[main node](302)[below right of=300]{$3$};
	\node[main node](303)[right of=302]{$4$};

	\draw[->] (100) -- (101) node[midway,above] {$1$};
	\draw[->] (102) -- (101) node[midway,above] {$2$};
	\draw[->] (103) -- (101) node[midway,right] {$3$};
	\draw[->] (103) -- (104) node[midway,above] {$4$};

	\path[every node/.style={font=\sffamily\small}]
	(200) edge node {} (201)
	(201) edge node {} (202)
	(200) edge node {} (202);

	\path[every node/.style={font=\sffamily\small}]
	(302) edge node {} (303);

	\draw[->] (150) -- (151) node[midway,above] {$L^i$};

	\draw[->] (150) -- (251) node[midway,above right=-.1cm] {$L^o$};
	
\end{tikzpicture}

\caption{The operation of the additional line graph operators, $L^i$ and $L^o$.} \label{fig-LiLo example}
\end{figure}
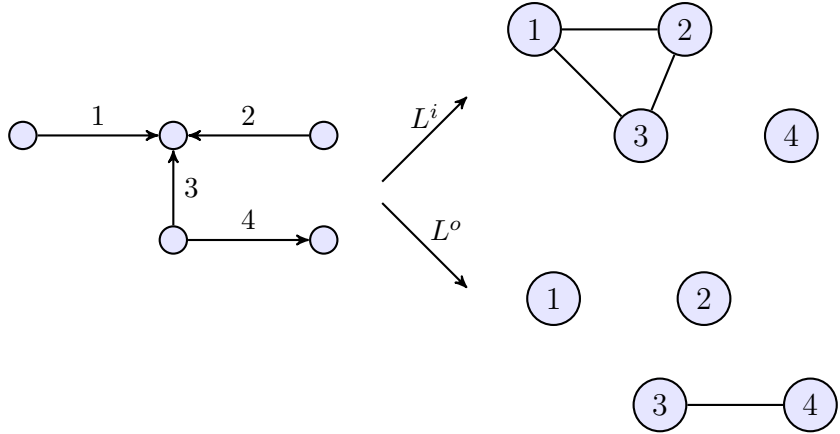

In order to write expressions for $C^{A,B}$ for a general tree, we need to introduce two more line graph operators. For a directed graph, the standard definition of its line graph is that two edges are connected if they share a vertex and one edge is oriented towards the vertex, and the other away from it. We introduce $L^i$ which connects edges if they share a vertex and are both directed \emph{towards} it, and $L^o$ which connects edges if they share a vertex and are both directed \emph{away from} it. This is illustrated by an example in figure \ref{fig-LiLo example}.

With these in hand we can write
\begin{align}
C^B &\propto C^o = I_{N-1} + A(L^o_G),\\
C^A &\propto b_L C^i - b_R C^o,
\end{align}
and %we now need to consider the action of $A(L^{i,o}(G))$ on the eigenvectors of $K$, i.e. on the eigenvectors of $A(L_G)$.
given a theory graph, one can now use these expressions and \eqref{Ctilde} to determine the largest Wilson coefficient, and hence the effective strong coupling scale. 

From \eqref{Ctilde} we see that due to the inverse powers of $\lambda(K(G))$, the largest coefficient will likely come from a term in which two of the fields are that with the minimal value of $\lambda(K(G))$, and so we isolate this dependence by writing
\be
\tilde{C}_\text{max} = \frac{f(G)}{\lambdamin(K(G))}, \label{Cmax}
\ee
where
\be
f(G) = \lambdamin(K(G))\, \mathrm{max} \left( \frac{1}{\sqrt{\lambda_i \lambda_j \lambda_k}} \left|\sum_{l m} U_{l i} U_{l j} C_{l m} U_{m k} \right| \right).
\ee
One may wonder why a third factor of $\lambdamin(K(G))^{-1/2}$ is not extracted, and the reason is twofold: whilst $C^{A,B}$ depends on $L^{i,o}_G$, and not just $L_G$, the similarities are still such that $\sum_{m} C_{lm} U_{m k} \sim \lambda_{k}$. 
This means that $\tilde C$ will not in general scale inversely with three powers of $\lambda(K(G))$ and so the maximum coefficient will not generally have \emph{three} fields with minimal $\lambda$; secondly the form \eqref{Cmax} will prove to have a nice interpretation in terms of graviton mass eigenstates, as explained in section \ref{sec-graviton masses}.

Having placed bounds on $\lambdamin(K(G))$ it remains to examine the behaviour of $f(G)$. 
Unfortunately, for given diameter, it is more difficult to determine which graphs extremise this, and in fact the dependence of those extreme values on the diameter is much weaker than for $\lambdamin(K(G))$. (The same is true if one considers its dependence on the maximum degree of the graph, for example.)

None-the-less we can say something about its expected value (taking the prior distribution to be uniform on the set of unlabelled trees of a given number of vertices); we find
\be
\langle f(T) \rangle \approx \frac{1}{\sqrt{d-1}}. \label{f approx}
\ee
Thus, given \eqref{lambda bounds} one has for $d>2$ (note that $d=2$ must be star graph, for which $\Lambda = \Lambda_3$)
\be
1.5 \left( \frac{\sqrt{d-1}}{N(d-2)} \right)^{1/3} \lesssim \frac{\Lambda}{\Lambda_3} \lesssim 2.1 \left( \frac{\sqrt{d-1}}{(d+1)^2} \right)^{1/3},
\ee
or for large $d$:
\be
\frac{1.5}{\sqrt{d}} \left( \frac{d}{N} \right)^{1/3} \lesssim \frac{\Lambda}{\Lambda_3} \lesssim \frac{2.1}{\sqrt{d}}. \label{Lambda bounds}
\ee
Although these bounds are not strict (since we have only taken the average value of $f(T)$) they are fairly inclusive. Only $2.5 \%$ (resp. $2.0 \%$) of trees with 15 (resp. 20) vertices exceed the upper bound, and $0.4 \%$ (resp. $0.5 \%$) fall below the lower bound.

\subsubsection{Higher order interaction terms} \label{sec-higherorder}

Having focussed only on the cubic scalar interactions we now briefly consider higher order (in the fields) interactions and explain why we would not expect them to lead to a lower effective cutoff than for cubic terms.

An extra scalar has to come in with two extra derivatives (because of the multi-gravity $\Lambda_3$ decoupling limit structure \cite{Noller:2015eda}), and hence an $(n+2)^\text{th}$ order term would look like
\be
\frac{1}{\Lambda_3^{3n}} \left( \frac{1}{\sqrt{ \lambda_{i_1} \cdots \lambda_{i_{n+2}} }} \sum_{lm} U_{li_{n+2}} \cdots U_{l i_2} C_{lm} U_{m i_1}  \right) \mathcal{L} ( \chi_{i_1}, \cdots, \chi_{i_{n+2}} ).
\ee
As before extract $n+1$ factors of $\sqrt{\lambdamin}$, and writing the sum that is left as $f_n$ we have that the effective cutoff is
\be
\frac{\Lambda}{\Lambda_3} = \frac{\lambdamin^{\frac{n+1}{6n}}}{f_n^{\frac{1}{3n}}}.
\ee
Since $\lambdamin$ is raised to an increasingly small power, $f_n$ would have to increase with $n$ to compensate (e.g. for a path graph, $\Lambda \sim 1/\sqrt{N}$ requires $f_n \sim N^{\frac{n}{2} -1}$). This seems unlikely, especially as naive power counting in $N$ would suggest opposite behaviour, and so it is reasonable to conclude that the cubic order terms give the lowest effective cutoff.

\subsection{Mass diagonalisation } \label{sec-mass}

Thus far we have ignored the fact that the \St scalars are mixed not only in their kinetic terms, but also in their mass terms, and really to find the propagating modes we must find the mass eigenstates. In practice we find that not including mass diagonalisation does not greatly affect the value of the strong coupling scale, and in this section we briefly investigate why this is.

Like their kinetic terms, the \St scalars do not initially have mass terms, but gain them from the spin-2 mass terms when the field redefinition $\sim h \to h - \pi \eta$ is applied. It turns out that their mass terms mix \St scalars which are \emph{two} links apart, rather than just adjacent, and so the simplest example is a tetra-metric path theory, $P_4$, since this requires three \St scalars. Taking the interaction terms to be the same, as described in the previous section, the spin-2 mass terms can be written,
\be
h_{\mn}^T
\begin{pmatrix}
m_{RR}^2 & m_{RL}^2 & 0 & 0 \\
m_{RL}^2 & 2m_{LL}^2 & m_{RL}^2 & 0 \\
0 & m_{RL}^2 & 2m_{RR}^2 & m_{RL}^2 \\
0 & 0 & m_{RL}^2 & m_{LL}^2
\end{pmatrix}
h_{\alpha\beta} M^{\mn\alpha\beta}, \label{tetra h mass term}
\ee
where $M^{\mn\alpha\beta}$ contains the information about the tensor structure of these terms and would be $\eta^{\mn\alpha\beta}$ for Fierz-Pauli structure, whereas the other matrix of constant coefficients details the `flavour' structure. Once the appropriate field redefinitions to remove the quadratic scalar-tensor mixing are applied, the scalar mass terms look like
\be
\mathcal{L}_{\rho^2} \propto \rho^T
\begin{pmatrix}
m_{RR}^2 + 2m_{LL}^2 - 2m_{RL}^2 & 2(m_{LL}^2 - m_{RL}^2) & -m_{RL}^2 \\
2(m_{LL}^2 - m_{RL}^2) & 2(m_{RR}^2 + m_{LL}^2 - m_{RL}^2) & 2(m_{RR}^2 - m_{RL}^2) \\
-m_{RL}^2 & 2(m_{RR}^2 - m_{RL}^2) & 2m_{RR}^2 + m_{LL}^2 - 2m_{RL}^2
\end{pmatrix}
\rho. \label{scalar mass terms}
\ee

If we have $m_{RR}^2 = m_{LL}^2 = -m_{RL}^2$,\footnote{For example, this is the case if the interaction terms are those of the so-called `minimal model' for which $\mathcal{L}_\text{int}(g,f) = \sqrt{-g} \left( 3 - \mathrm{tr} \sqrt{g^{-1} f} + \det \sqrt{g^{-1} f} \right)$. In general the requirement is $\beta_0 + 4\beta_1 + 6\beta_2 + 4\beta_3 + \beta_4 = 0 = \beta_1 + 3\beta_2 + 3\beta_3 + \beta_4$, or equivalently to the absence at the perturbative level of constant and tadpole terms in the Lagrangian, whose presence at any rate would be equivalent to considering expanding around a different background.} then the mass matrix in \eqref{scalar mass terms} becomes proportional to
\be
\begin{pmatrix}
5 & 4 & 1 \\
4 & 6 & 4 \\
1 & 4 & 5
\end{pmatrix}
= 4 + 4 A(L_{P_4}) + A(L_{P_4})^2,
\ee
where we recall that $A$ is the adjacency matrix. It is easy to see that this property generalises to other trees, i.e.
\be
\mathcal{L}_{\rho^2} \propto \rho^T \left( 4 + 4 A(L_T) + A(L_T)^2 \right) \rho.
\ee
Now since the eigenvectors of $K(T)$ are the eigenvectors of $A(L_T)$, and hence of $A(L_T)^2$, the mass diagonalistion becomes trivial in this case, being already performed by the kinetic diagonalisation.

When $m_{RR}^2 = m_{LL}^2 = -m_{RL}^2$ is not satisfied by the interaction terms, then it is no longer possible to write the scalar mass matrix in terms of $A(L_T)$ and powers thereof, but, especially if $m_{LL}^2 \sim m_{RR}^2$, its structure is still very similar to that of $A(L_T)$ and its powers, and so it is perhaps unsurprising that it is safe to neglect its diagonalisation, since it will already be approximately diagonalised by the kinetic term diagonalisation procedure.

\section{Relation to graviton masses} \label{sec-graviton masses}

In the case of a path graph, one can write the strong coupling scale as \cite{deRham:2013awa}
\be
\Lambda_\text{eff} = \left( m_1^2 M_\text{Pl} \right)^{1/3}, \label{Lambda eff}
\ee
where $m_1$ is the mass of the lightest massive graviton state, and $M_\text{Pl}$ is the physical Planck mass, which are respectively related to the mass parameters $m$ and $M$ entering the action. Intuitively this makes sense, since if there is a gap between the lightest massive graviton and the next, then at energies below the mass of the next to lightest graviton the theory is effectively bi-gravity and so the strong coupling scale should be $\Lambda_3$ built out of mass parameters $m_1$ and $M_\text{Pl}$. In this section we investigate to what extent this is true in general.

First we need to find the graviton mass spectrum for a generic theory; from \eqref{tetra h mass term} one can see that if $m_{LL}^2 = m_{RR}^2 = -m_{RL}^2 = m^2$, extending to a general graph the mass matrix satisfies
\be
M(G) = m^2 L(G) = m^2 \left( D(G) - A(G) \right),
\ee
where $D(G)$ is the diagonal matrix whose $ii$ entry is the degree of vertex $i$. The combination $L(G)$, of this and the adjacency matrix, is called the \emph{Laplacian} (or \emph{Kirchhoff}) matrix of the graph and its properties have been widely studied.\footnote{When $m_{LL}^2 = m_{RR}^2 = -m_{RL}^2$ is not the case, one would not expect that all the conclusions of this section would remain exactly correct, but again, due to the similarity, even in the general case, of the structure of $M(G)$ to $L(G)$, one would expect the general conclusions to remain accurate.}

We note some basic facts about $L(G)$ \cite{brualdimutually} and their interpretation in terms of multi-gravity theories.
\begin{itemize}
\item $L(G)$ has one zero eigenvalue, i.e. there is one massless graviton, since there is one unbroken copy of $GC$.\footnote{For graphs with multiple separate connected components, there is a corresponding number of zero eigenvalues.}
\item The eigenvector corresponding to the zero eigenvalue is $(1, \cdots, 1)^T$, i.e. the sum mode $\sum_i h_{\mu\nu}^{(i)}$ is always massless.
\item $L(G)$ is positive semi-definite, i.e. none of the gravitons are tachyonic.
\item The smallest non-zero eigenvalue of $L(G)$ is called the algebraic connectivity, and is larger for graphs which are more ``connected," i.e. more ``connected" theories have a larger mass gap.
\end{itemize}

\subsection{Relation to $K(G)$}

Given a graph $G$, and an arbitrary orientation of its edges, the \emph{incidence matrix} of a graph, is the $|E| \times |V|$ matrix with elements
\be
B(G)_{ij} =
\begin{cases}
1 & \text{if edge }i\text{ \emph{leaves} vertex }j \\
-1 & \text{if edge }i\text{ \emph{arrives} at vertex }j \\
0 & \text{otherwise}
\end{cases},
\ee
then one has
\be
L(G) = B^T B, \qquad \text{and} \qquad K(G) = B_+ B_+^T,
\ee
where $B_{+,ij} = |B_{ij}|$. Now for a tree (or, indeed, any graph which is odd-cycle-free) one can consistently orient the edges such that each vertex has edges either all leaving, or all arriving, in which case one has $B_{ij} = (-1)^{\sigma_j} B_{+,ij}$, where $\sigma_j = 0, 1$ appropriately. 

Thus for a tree one has $K(T) = B B^T$, and hence the non-zero eigenvalues of $L(T)$ and $K(T)$ are identical, i.e. $\mathrm{spec}(L(T)) = \{ 0 \} \cup \mathrm{spec}(K(T))$, where $\mathrm{spec}(X)$ is the spectrum of a matrix $X$. In particular this means that $\lambdamin(K(T)) = m_1^2$, and hence from \eqref{Cmax} one has
\be
\Lambda_\text{eff}^3 = m_1^2 \frac{M}{f(T)},
\ee
and \eqref{Lambda eff} is recovered if $f(T) = M/M_\text{Pl}$. Thus in order to consider whether \eqref{Lambda eff} gives an accurate expression for the strong coupling scale one needs an expression for the physical Planck mass to compare with $M/f(G)$; we now consider two approaches to this.

\subsection{Dimensional deconstruction}

A path or cycle graph has a natural interpretation as a discretised extra dimension \cite{ArkaniHamed:2001ca, ArkaniHamed:2003vb, Schwartz:2003vj, Deffayet:2003zk, deRham:2013awa, Kan:2002rp}, which leads to  a relation between $M$ and $M_\text{Pl}$. The number of sites in the discretised theory is $N = m R$, where $R$ is the size of the extra dimension, and hence the most massive graviton has mass $m_N = 2 m \sim N/R$. One also has $M_\text{Pl}^2 = M^3 R$, and hence requiring the validity of the effective theory via $m_N = M$, one has the relation $M_\text{Pl}^2 \sim N M^2$ \cite{deRham:2013awa}, which indeed agrees with $M_\text{Pl} = M/f(P_N)$.

More complicated graphs do not have so obvious an interpretation in terms of higher dimensional theories, however using that $M_\text{Pl}^2 = M^3 R$ is still expected to hold for general dimensionally reduced theories and $m_N = M$, one has
\be
f(T) = \frac{M}{M_\text{Pl}} = \frac{1}{\sqrt{m_N R}}, \label{f higher dim}
\ee
and there are several ways to proceed: one could postulate a relation between $R$ and the graph structure, and see whether \eqref{f higher dim} is satisfied, or alternatively one could take \eqref{f higher dim} as a definition of $R$.

As an example, consider the star graph: as already mentioned, $f({S_N}) = 1$, and we also have $m_N^2 = N m^2$, and hence $R = m^{-1}/\sqrt{N}$. On its face this is a little peculiar: keeping the `discretisation scale,' $m^{-1}$, fixed, as we increase the number of sites, the size of the extra dimension being described actually decreases. In reality though we probably should not expect \eqref{Lambda eff} to give a good expression for the strong-coupling scale in this case, because the graviton mass spectrum is highly degenerate:
\be
\mathrm{spec}(L({S_N})) = \{0, 1^{N-2}, N\}.
\ee
i.e. there are $N-2$ gravitons with mass $m_1 = m$. Thus it is no longer true that at low enough energies one can treat the theory as simply bi-gravity, and we have no reason to expect the cutoff to be \eqref{Lambda eff}, with the attendant interpretation as resulting from a higher dimensional theory. This is not to say that a star graph theory has no place within dimensional deconstruction, merely that its interpretation is a little more tricky.

Denoting by $\Delta$ the largest degree of any vertex in $T$, one has the following bounds \cite{doi:10.1137/S0895480191222653, Stevanović200335}
\be
\Delta+1 \leq m_N^2 < \Delta + 2 \sqrt{\Delta - 1},
\ee
and hence from \eqref{f approx}, $R \sim m^{-1} d/\sqrt{\Delta}$. Intuitively this makes sense since as one increases $\Delta$ (resp. $d$) whilst keeping $d$ (resp. $\Delta$) fixed, it is increasingly similar to $K_{1,\Delta-1}$ (resp. $P_{d+1}$). Thus a theory will avoid the counter-intuitive behaviour described above as its theory graph `grows' if $d$ grows more quickly than does $\sqrt{\Delta}$.

\begin{figure}[tp]
\centering
\begin{tikzpicture}[-,>=stealth',shorten >=0pt,auto,node distance=2cm,
  thick,main node/.style={circle,fill=blue!10,draw,font=\sffamily\large\bfseries},arrow line/.style={thick,-},barrow line/.style={thick,->},no node/.style={plain},rect node/.style={rectangle,fill=blue!10,draw,font=\sffamily\large\bfseries},red node/.style={rectangle,fill=red!10,draw,font=\sffamily\large\bfseries},green node/.style={circle,fill=green!20,draw,font=\sffamily\large\bfseries},yellow node/.style={rectangle,fill=yellow!20,draw,font=\sffamily\large\bfseries}]

	\node[main node](100){};
	\node[main node](101)[right of=100]{};
	\node[draw=none,fill=none](102)[right=1cm of 101]{};
	\node[draw=none,fill=none](103)[right=1cm of 102]{};
	\node[main node](104)[right=1cm  of 103]{};

	\node[main node](99)[above left of=100]{};
	\node[main node](98)[below left of=100]{};

	\path[every node/.style={font=\sffamily\small}]
	(100) edge node {$1$} (101)
	(101) edge node {$2$} (102)
	(103) edge node {$d-1$} (104)
	(100) edge node {} (99)
	(100) edge node {} (98);

	\draw[-,dashed] (102) to (103);

	\draw[-,dashed] ($(99)+(0.1,-0.5)$) arc (135:225:1.3);
	\draw(100) -- (158.5:1.5);
	\draw(100) -- (180:1.5);
	\draw(100) -- (202.5:1.5);

	\node[draw=none,fill=none](1)[left=1.5cm of 100]{$\Delta-1$};

\end{tikzpicture}

\caption{The `mace graph,' $M_{d,\Delta}$.}\label{fig-mace graph}
\end{figure}
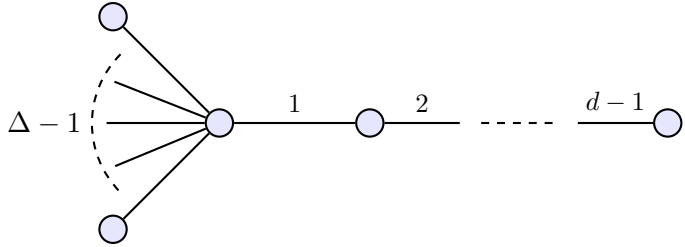

As a further example, consider the `mace graph,' $M_{d,\Delta}$, which we define in figure \ref{fig-mace graph}; this indeed has $R \sim d / \sqrt{\Delta}$. Curiously however, the fact that $R$ decreases when $\Delta$ grows quickly enough compared to $d$ cannot in this case be blamed on there being more than one graviton at, or close to, mass $m_1$, as in the case of the star graph. We show in appendix \ref{app-dumbbell graph} that for $d>4$,
\be
\frac{m_2}{m_1} > \frac{3}{2}
\ee
(and in fact numerically we find $m_2 > 2 m_1$), i.e. the separation between the first two massive gravitons is always of order the mass gap of the theory, which is precisely the case for the path graph theory (which has $m_2 \approx 2 m_1$.)

\subsection{Coupling to matter}

Outside of the framework of dimensional deconstruction an expression for the physical Planck mass can be derived by coupling the theory to matter, since this coupling strength is what determines the strength of gravity, and hence $M_\text{Pl}$. Considering this question in the detail it deserves is unfortunately beyond the scope of this work, except to make a few elementary observations.

Obviously in general this will depend on how one couples to matter (the known consistent non-derivative such couplings are those of \cite{deRham:2014naa,Noller:2014sta,Melville:2015dba}), for instance if one couples to just one site, then this will clearly depend on at least the graph structure in the vicinity of that vertex.

A more realistic scenario might be that one wants to couple to the (perturbative) massless mode. Regardless of graph structure, the massless mode is $h_{\mn}^{(m=0)} = \sum_i h_{\mn}^{(i)}$; this leads to a kinetic term $N M^2 h^{(m=0)} \mathcal{E} h^{(m=0)}$, and hence canonical normalisation involves $h^{(m=0)} \to \frac{1}{\sqrt{N} M} h^{(m=0)}$. At the non-linear level coupling to this could arise from coupling to an effective vielbein \cite{Noller:2014sta} $E^{\text{eff}} = \sum_i E^{(i)}$; linearising, and then canonically normalising one has
\be
\frac{1}{\sqrt{N} M} \int d^4 x\, h_{\mn}^{(m=0)} T^{\mn},
\ee
and thus the coupling strength is $M_\text{Pl} = \sqrt{N} M$, i.e. the same as in dimensional deconstruction using a path graph.
This gives $f = M/M_{\rm Pl} \sim 1/\sqrt{N}$, which is satisfied for trees with $d \sim N$ (see section \ref{sec-bounds}). However there are many trees which violate this, and hence violate \eqref{Lambda eff}; furthermore not all of these are ones which have a degenerate lightest graviton mass (for which we already do not expect \eqref{Lambda eff} to hold). 
This would seem to advise against using \eqref{Lambda eff} in general to estimate the strong coupling (at least when one is coupling matter to the massless mode).\footnote{Of course, this relies on starting off with order one parameters, and in particular coupling to $E^{\text{eff}} = \sum_i E^{(i)}$, rather than, say $E^{\text{eff}} = \frac{1}{\sqrt{N}} \sum_i E^{(i)}$ -- though if the latter were chosen one still would not have a result which was valid for all graphs.}

\section{Conclusions}\label{sec-conc}

In this paper we have considered the question of how the strong coupling scale of a multi-gravity theory is related to its topology, i.e. to the presence and structure of interactions between the different fields. We have shown that this can elegantly be rephrased in terms of various properties of the corresponding theory graph. In particular we have placed bounds \eqref{Lambda bounds} on the strong coupling scale. The parameter to which these bounds are most sensitive is the diameter of the graph. One would expect that tighter bounds could in principle be derived by taking into account further graph properties (of course at the expense of complexity and/or generality). These bounds give us a sense of how the range of validity of the theory behaves as one changes the graph and hence the theory. One important finding is that the strong coupling scale of a multi-gravity theory can never be raised simply by adding extra spin-2 fields and/or interactions between them.

We have also considered when the expression \eqref{Lambda eff} ($(m_1^2 M_\text{Pl})^{1/3}$) for the strong coupling scale, which one gets by just considering the lightest massive graviton, is valid. The lightest graviton mass turns out to be the algebraic connectivity of the graph, and fits easily with the previously computed bounds. One also has to determine the physical Planck mass, and we have considered doing this via appealing to a higher-dimensional interpretation of the theory, and by considering the coupling of the theory to matter. 

The latter approach implies that \eqref{Lambda eff} is not universally valid, whilst the former yields curious behaviour in which, for a fixed discretisation scale, the size of the nominal extra dimension decreases as the size of the graph increases. Whilst we would not expect theories which have more than one graviton at mass scale $m_1$ to have a cutoff described by \eqref{Lambda eff}, interestingly it also does not hold in some theories which do have a separation between the first two massive gravitons.

It would be useful to further investigate the question of whether and how general theory graphs might be related to higher dimensional theories, in the way that path and cycle graph theories can. Our results here seem to indicate that the usual approach to interpretation is not appropriate.
Finally, we have only considered bi-metric interaction terms, whilst using vierbeins one may write interaction terms which involve more than two fields. It would be interesting to extend this work to those theories.
\\
\\
\noindent {\bf Acknowledgements: } We would like to thank Claudia de Rham and Andrew Tolley for several stimulating discussions regarding this work. JHCS is supported by STFC. JN acknowledges support from the Royal Commission for the Exhibition of 1851 and BIPAC.

\appendix

\section{The spectrum of the generalised barbell graph} \label{app-dumbbell graph}

The line graph of the graph depicted in figure \ref{fig-lower bound graph} consists of two complete graphs of $\lfloor \frac{N+3-d}{2} \rfloor$ and $\lceil \frac{N+3-d}{2} \rceil$ vertices connected by a path graph with $d-3$ edges. This is an example of a `generalised barbell graph', $B_{a,b,c}$, which we define to consist of a path $P_{b+2}$ which has each endpoint in common with each of two complete graphs, $K_a$ and $K_c$.\footnote{In this notation the ordinary barbell graph would be $B_{n,0,n}$.}

The adjacency matrix is
\be
A(B_{a,b,c}) =
\begin{pmatrix}
J_{a,a} & \begin{matrix} 0_{a-1,1} & 0_{a-1,b-1} \\ 1 & 0_{1,b-1} \end{matrix} & 0_{a,c} \\
\begin{matrix} 0_{1,a-1} & 1 \\ 0_{b-1,a-1} & 0_{b-1,1} \end{matrix} & T_b & \begin{matrix} 0_{b-1,1} & 0_{b-1,c-1} \\ 1 & 0_{1,c-1} \end{matrix} \\
0_{c,a} & \begin{matrix} 0_{1,b-1} & 1 \\ 0_{c-1,b-1} & 0_{c-1,1} \end{matrix} & J_{c,c}
\end{pmatrix} - I_{a+b+c},
\ee
where $J_{n,n}$ is the $n \times n$ matrix of 1's, and $T_n$ is a tri-diagonal matrix of 1's. The characteristic polynomial $\Delta_{a,b,c}$ obeys the following recurrence relation in $a$:
\be
\Delta_{a,b,c} = | \lambda I_{a+b+c} - A(B_{a,b,c}) | = \lambda \Delta_{a-1,b,c} + (a-2) \Delta^{(1)}_{a-2,b,c} + \Delta^{(2)}_{a-2,b,c},
\ee
where 
\be
\Delta^{(1)}_{a,b,c} = 
\left| \lambda I_{a+b+c+1} - A(B_{a+1,b,c}) - 
\begin{pmatrix}
\lambda + 1 & 0_{1,a+b+c} \\
0_{a+b+c,1} & 0_{a+b+c,a+b+c}
\end{pmatrix}\right|
= - \Delta_{a,b,c} + (a-1) \Delta^{(1)}_{a-1,b,c} + \Delta^{(2)}_{a-1,b,c},
\ee
and
\be
\Delta^{(2)}_{a,b,c} = 
\left| \lambda I_{a+b+c+1} - A(B_{a+1,b,c}) - 
\begin{pmatrix}
\lambda + 1 & 0_{1,a} & 1 & 0_{1,b-1+c} \\
0_{a+b+c,1} & 0_{a+b+c,a} & 0_{a+b+c,1} & 0_{a+b+c,b-1+c}
\end{pmatrix}\right|
= - (\lambda + 1)^a \Delta_{0,b,c}.
\ee
These recurrence relations can be solved, noting that $\Delta_{1,b,c} = \Delta_{0,b+1,c}$, to get
\be
\Delta_{a,b,c} = - (1 + \lambda)^{a-2} \left( (a-1) \Delta_{0,b,c} - (\lambda - (a-2)) \Delta_{0,b+1,c} \right). \label{Deltaa}
\ee

Now $\Delta_{0,b,c}$ obeys the following recurrence relation in $b$:
\be
\Delta_{0,b,c} = \lambda \Delta_{0,b-1,c} - \Delta_{0,b-2,c},
\ee
which, by using $\Delta_{0,0,c} = (\lambda + 1)^{c-1} (\lambda - (c-1))$ and $\Delta_{0,1,c} = \lambda \Delta_{0,0,c} - (\lambda + 1)^{c-2} (\lambda - (c-2))$, can be solved to give
\begin{align}
\Delta_{0,b,c} = &\frac{1}{2^{b+3}}(\lambda + 1)^{c-2} \left[ (c-1) \Bigg( \left(\lambda - \sqrt{\lambda^2 - 4}\right)^{b+2} + \left(\lambda + \sqrt{\lambda^2 - 4}\right)^{b+2} \right) \nonumber \\
&+ (\lambda(c-3) + 2(c-2)) \left( \frac{\left(\lambda - \sqrt{\lambda^2 - 4}\right)^{b+2} - \left(\lambda + \sqrt{\lambda^2 - 4}\right)^{b+2}}{\sqrt{\lambda^2 - 4}} \right) \Bigg]. \label{Delta0}
\end{align}

We finally change variables: $\lambda = -2\cos \theta$, and substitute \eqref{Delta0} in \eqref{Deltaa}, and a little algebra yields
\be
\Delta_{a,b,c} = (-1)^b (1-2 \cos \theta)^{a+b-4} \mathrm{cosec}\,\theta\, f_{a,b,c}(\theta),
\ee
where
\begin{align}
f_{a,b,c}(\theta) = &\sin((b+5)\theta) + (a+c-4) \sin((b+4)\theta) + \left( (a-3)(c-3) -1 \right) \sin((b+3)\theta) \nonumber \\
&- 2 (a-2)(c-2) \sin((b+2)\theta) + (a-2)(c-2) \sin((b+1)\theta). \label{fabc}
\end{align}
$f_{N,d}$ in \eqref{fNd} is then just \eqref{fabc} with $a = c = \frac{N+1-d}{2}$ and $b = d - 4$.

\subsection{Spectrum of $L(M_{d,\Delta})$}

One has $L_{M_{d,\Delta}} = B_{\Delta,d-4,2}$,\footnote{Note there is a degeneracy in the final two parameters: $B_{\Delta,d-4,2} = B_{\Delta,d-3,1} = B_{\Delta,d-2,0}$.} and hence 
\be
\mathrm{spec}(L(M_{d,\Delta})) = \{0, 2(1-\cos \theta^*_i), 1^{d+\Delta-8} \},
\ee
where $\theta^*_i$ is a root of
\begin{align}
f_{\Delta,d-4,2}(\theta) &= \sin((d+1)\theta) + (\Delta-2)(\sin(d \theta) - \sin((d-1)\theta) ) \\
&= (\Delta - 1) \sin(d \theta) \sin(\theta) \left[ \mathrm{cot} (d\theta) - \left( \frac{\Delta-3}{\Delta-1} \mathrm{cot}(\theta) - \frac{\Delta-2}{\Delta-1} \mathrm{cosec}(\theta) \right) \right].
\end{align}
From the square bracket we see that $(n-\frac{1}{2})\frac{\pi}{d} < \theta^*_n < n \frac{\pi}{d}$. Hence we have for $d>4$
\be
\frac{m_2}{m_1} \approx \frac{\theta^*_2}{\theta^*_1} > \frac{3}{2}.
\ee

\bibliographystyle{JHEP}
\bibliography{SCinMultiG-bib}

\end{document}